\documentclass[sigconf]{acmart}
\usepackage{appendix}
\usepackage{booktabs}
\usepackage{multirow}
\usepackage{makecell}
\usepackage{verbatimbox}
\usepackage{subfigure}
\usepackage{graphicx}
\usepackage{xspace}
\usepackage{bm}
\usepackage{mathtools}
\newcommand{\etal}{\emph{et~al.}\xspace}
\newcommand{\eg}{\emph{e.g.},\xspace}
\newcommand{\ie}{\emph{i.e.},\xspace}

\newcommand\figref[1]{Fig.~\ref{#1}}

\newcommand\tabref[1]{Table~\ref{#1}}

\newcommand\secref[1]{Sec.~\ref{#1}}

\newcommand{\fakeparagraph}[1]{\vspace{1mm}\noindent\textbf{#1.}}

\newcommand{\sysname}{ KADM\xspace }

\AtBeginDocument{%
  \providecommand\BibTeX{{%
    \normalfont B\kern-0.5em{\scshape i\kern-0.25em b}\kern-0.8em\TeX}}}

\acmSubmissionID{fp0151}

\copyrightyear{2022}
\acmYear{2022}
\setcopyright{acmcopyright}\acmConference[SIGIR '22]{Proceedings of the 45th
International ACM SIGIR Conference on Research and Development in Information
Retrieval}{July 11--15, 2022}{Madrid, Spain}
\acmBooktitle{Proceedings of the 45th International ACM SIGIR Conference on
Research and Development in Information Retrieval (SIGIR '22), July 11--15, 2022,
Madrid, Spain}
\acmPrice{15.00}
\acmDOI{10.1145/3477495.3532070}
\acmISBN{978-1-4503-8732-3/22/07}
\begin{document}
% \fancyhead{}
%\title{Jointly Learn Local and Global Information Representations for Top-$N$ Recommendation}
\title{Unify Local and Global Information for Top-$N$  Recommendation}
\author{Xiaoming Liu, ~Shaocong Wu, ~Zhaohan Zhang, and ~Chao Shen}
%\authornote{Corresponding author.}
\affiliation{%
  \institution{Ministry of Education Key Lab for Intelligent Networks and Network Security, \\School of Cyber Science and Engineering,  Xi'an Jiaotong University}
  \streetaddress{}
  \city{No.28, Xianning West Road, Xi'an}
  \state{Shaanxi}
  \country{China}
  }
  \email{{xm.liu,chaoshen}@xjtu.edu.cn,  {shaocong.wsc,zzh1103}@stu.xjtu.edu.cn}


\begin{abstract}
Knowledge graph (KG), integrating complex information and containing rich semantics, is widely considered as side information to enhance the recommendation systems.
However, most of the existing KG-based methods concentrate on encoding the structural information in the graph, without utilizing the collaborative signals in user-item interaction data, which are important for understanding user preferences. 
Therefore, the representations learned by these models are insufficient for representing semantic information of users and items in the recommendation environment. 
The combination of both kinds of data provides a good chance to solve this problem, but it faces the following challenges: 
i) the inner correlations in user-item interaction data are difficult to capture from one side of the user or item; 
ii) capturing the knowledge associations on the whole KG would introduce noises and variously influence the recommendation results;
iii) the semantic gap between both kinds of data is hard to alleviate.

To tackle this research gap, we propose a novel duet representation learning framework named \sysname to fuse local information (user-item interaction data) and global information (external knowledge graph) for the top-$N$ recommendation, which is composed of two separate sub-models. 
One learns the local representations by discovering the inner correlations in local information with a knowledge-aware co-attention mechanism, 
and another learns the global representations by encoding the knowledge associations in global information with a relation-aware attention network.
The two sub-models are jointly trained as part of the semantic fusion network to compute the user preferences, which discriminates the contribution of the two sub-models under the special context.
We conduct experiments on two real-world datasets, 
and the evaluations show that \sysname significantly outperforms state-of-art methods.
Further ablation studies confirm that the duet architecture performs significantly better than either sub-model on the recommendation tasks.

\end{abstract}


% <concept_id>10002951.10003260.10003261.10003269</concept_id>
% <concept_desc>Information systems~Collaborative filtering</concept_desc>

% \ccsdesc[500]{Information systems~Collaborative filtering}



\maketitle

\section{Introduction}

With the rapid development of Internet technology, the amount of online data has increased sharply. 
The massive information would overwhelm users so that it is time-consuming for them to filter out their favorite information among a large number of choices.
To alleviate this effect, recommendation systems have become a vital and indispensable tool to assist users in making decisions.

\fakeparagraph{Prior Works and Limitations}
The recommendation system attracts intensive research interest and derives broad applications \cite{sarwar2001item}.
Conventional collaborative filtering (CF) methods \cite{he2017neural, weimer2008adaptive, zhang2016discrete}, which provide recommendation based on the user-item interaction data, have made a significant success. 
However, CF-based methods usually suffer from the data sparsity and cold-start issues \cite{wang2018tem} due to the fact that even the most active users just have interacted with a small percentage of items in the recommendation environment.
Meanwhile, most of them are unable to thoroughly encode  the collaborative signals in interaction data, because they can only capture one-side influence from users or items and model the shallow correlations between users and items.
To solve these issues, many methods try to exploit different types of side information (\eg item description \cite{chen2012svdfeature}, user profile \cite{guo2019exploiting} and social network \cite{jamali2010matrix}). 
For instance, a number of trust-aware recommendation methods \cite{ma2011recommender, guo2012simple} are proposed based on the assumption that users may share similar preferences with their trusted users.

Recently, with the development of the semantic web, introducing knowledge graphs (KGs) into recommendation systems as side information has attracted extensive attention. 
In contrast with other forms of side information (\eg social network), which are generally limited to capturing features with homogeneous information, KGs are heterogeneous graphs connecting various types of features related to users or items in a unified global representation space \cite{sun2019research}. 
The structural information in KGs helps to explore the potential connections between users or items from different perspectives, which is beneficial for improving the performance of recommendation  algorithms \cite{luo2014hete, catherine2016personalized, sun2018recurrent, wang2021learning, xu2022ckgat}.
However, existing KG-based methods still share several common limitations:
First, most methods focus on knowledge associations in KG without exploiting the user-item interaction data, which are insufficient to represent user preferences;
Second, most studies conduct information propagation on the whole graph, which may introduce negative noise from irrelevant entities;
Thirdly, most works equally treat information from different relation-paths, which is contrary to the real recommendation scenarios.

\begin{figure}[!t]
  \centering
  \includegraphics[width=0.4\textwidth]{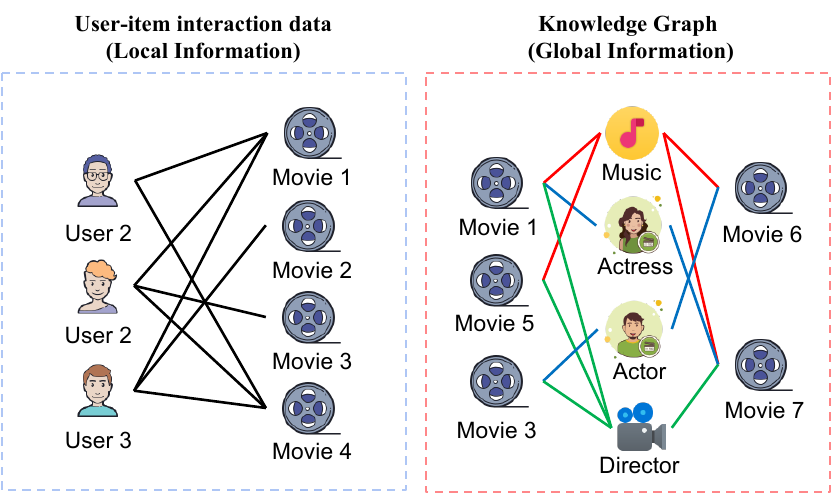}
  \caption{Illustration of local Information and global information. Local information is the user-item interaction data, which is a bipartite graph with users and items and interactions. Global information is a heterogeneous knowledge graph with multiple types of relationships and entities.}
  \label{fig:information}
%\vspace{-0.35 in}
\end{figure}

\fakeparagraph{Motivations and Relations} 
Comprehensive analysis reveals that user-item interaction data and KG could indicate the people's decision patterns from different perspectives.
Concretely, the collaborative signals in user-item interaction could introduce two kinds of effects from both item and user levels.
For item level, users tend to buy items similar to their historical items, which reflects historical unchanged and independent user preferences.
For another, people may be more likely to choose such item that is frequently purchased by other users together with the item they like. 
Moreover, there are rich knowledge associations in KG, which could indicate semantic relations between users or items from the attribute level.
For example, if two movies have the same director or actor, a user who has watched  one may be also willing to watch the other.
Furthermore, there are multiple kinds of relation-paths in KG.
The different relation-paths would have varying degrees of impact on user preferences. 
For example, a user is more likely to choose movies from the perspective of movie casting rather than the director.

In this paper, we respectively define the user-item interaction data and knowledge graph as the local and global information according to their characteristics and different impacts for the recommendation, as illustrated in  \figref{fig:information}.
Following the above analysis, the two-fold information can be combined and complemented each other to jointly infer a user's decision on the target item, but with the following challenges:
1) How to discover the inner correlations between users and items in local information?
2) How to eliminate the negative effect of noise triplets and model the influence of different relation-paths in global information?
3) How to alleviate the semantic gap between local and global information?

\fakeparagraph{Methodologies and Results}
To address the above challenges, we propose a \textbf{K}nowledge-\textbf{A}ware \textbf{D}uet \textbf{M}odel (named \sysname) comprised of two sub-models to respectively formulate user preferences based on local and global information:
i) Local model, which computes the local representation of each user or item by encoding the collaborative signals in local information.
It firstly represents each user and item with their collaborative neighbors.
Secondly, it captures the inner correlations by calculating the attention matrix based on the collaborative neighbors and further obtains the attention vectors with pooling operation. Finally, it could generate the local representations of users and items by attentively aggregating adjacent neighbors’ representation that reflects users’ preferences from both user and item side.
ii) Global model, which learns the global representation of each user or item by capturing the rich knowledge associations in global information.
It firstly extracts the enclosing subgraph of the user-item pair from global information.
Secondly, it attentively aggregates the semantic information propagated by different relation-paths in subgraph to update the representations of nodes.
Finally, it could generate the global representations of users and items by combining the aggregated features  with their own features.

Then, since local information as well as global information could complement each other to jointly influence a user's decision on target item and the importance of them would vary for distinct user-item pairs, we utilize a gating network to dynamically weigh the two sub-models, which has been proven useful to control importance of different information flows in the network, \ie the two sub-models are jointly trained as part of the gating network.

The major contributions in this paper are summarized as follows:
\begin{itemize}
\item \textbf{General Concepts:} 
We distinguish the concepts of the user-item interaction data from KG according to their characteristics and different impacts for the recommendation. 
Then we respectively define them as local and global information.

\item \textbf{Duet Recommendation Architecture:} We propose a novel duet architecture  model for top-$N$ recommendation, which takes advantage of
both global and local views to investigate user preferences.
In the duet recommendation architecture, local model takes the knowledge-aware co-attention mechanism to discover the inner correlations by encoding the collaborative signals in local information, and global model utilizes the relation-aware GNN to capture the knowledge associations in the enclosing subgraph extracted from global information.

\item \textbf{Gating network Semantic Fusion:} To eliminate the semantic gap between the two types of information, the gating network, based on a linear unit, dynamically weighs different impacts from two information according to specific contexts.

\item \textbf{Outstanding Performance:} 
We deploy \sysname on two real-world datasets.
The experiment results demonstrate the state-of-the-art performance of \sysname, the effectiveness of the proposed components, as well as its possible interpretability for modeling user preferences.

\end{itemize}

\section{Related work}
This section reviews the related works that are relevant to our work.

\fakeparagraph{CF-based recommendation}
Collaborative filtering (CF) is a technique widely used in recommender systems, which leverages the user-item feedback data to model the user preference. 
It mainly consists of neighbor-based methods \cite{sarwar2001item, koren2008factorization} and matrix factorization (MF) methods \cite{koren2009matrix, mnih2008probabilistic}. 
And recently, many CF approaches are combined with new deep learning techniques \cite{dziugaite2015neural, ying2018graph}.
However, although these methods could sometimes achieve good recommendation performance, most of them still suffer from data sparsity and cold-start problems, and can only model shallow relationships between users and items.

\fakeparagraph{KG-based recommendation}
Knowledge graph (KG) is widely used as auxiliary information to enhance recommendation systems, and has achieved effective results. 
It mainly consists of three categories, \ie embedding-based methods \cite{zhang2016collaborative,wang2018dkn,huang2018improving}, path-based methods \cite{guo2020survey,hu2018leveraging,shi2015semantic} and propagation-based methods \cite{wang2018ripplenet,wang2019knowledge,zhao2019intentgc}.
Among them, the propagation-based methods usually could achieve state-of-the-art results by recursively propagating the information from multi-hop nodes to refine the representation of users and items over the entire KG.
For instance, 
Wang \etal \cite{wang2019kgat} propose the concept of collaborative knowledge graph (CKG) to encode user behaviors and item knowledge as a unified relational graph, and further explore high-order connectivity with semantic relations in CKG for the recommendation.
MVIN \cite{tai2020mvin} gathers personalized knowledge information in the KG (user view) and further considers the difference among layers (entity view) to ultimately enhance item representations.
CKAN \cite{wang2020ckan} explicitly encodes the user-item interactions and naturally combines them with knowledge associations in an end-to-end manner.

Compared to previous methods, there are several key advantages in our proposed model: 
i)  \sysname combines the collaborative signals from the user-item interaction data with knowledge associations in a dual structure. 
Both kinds of information could complement each other to achieve better recommendation performance.
ii) With relation-aware attention mechanism, \sysname captures the various influence from different relation-paths during the information propagation process, while most of the previous methods are node-based and treat them equally.
iii) 
\sysname is based on the computation of enclosing subgraph extracted from the specific context while most methods work directly on the whole graph. 
It can alleviate the negative effects of irrelevant nodes in the propagation process and reduce the size of the graph to save computing resources.

\fakeparagraph{Dual Mechanism}
There are many dual phenomenons in real-life which inspire several dual structures in model design.
For instance, Xia \etal \cite{xia2018model} propose a model-level dual learning framework to merge the training of two dual tasks. 
DGCN \cite{zhuang2018dual} extends GCN to dual structures, jointly considering both the local and global consistencies in the graph. 
Cheng \etal \cite{cheng2018delf} propose a novel deep latent factor model named DELF with dual embeddings of users and items for recommendation.
DANSER \cite{wu2019dual} includes two dual graph attention networks to learn deep representations for social effects in recommendation systems.

Compared to previous methods, our model possesses several key differences: 
i)\sysname designs the specific sub-model for each type of data, while others utilize sub-models of symmetric structures.
ii) In the semantic fusion module, \sysname dynamically weighs the importance of the representations from different sub-models, while others usually straightforwardly concatenate different features vectors.

\section{PROBLEM FORMULATION}
Assume  that there are $m$ users and $n$ items, we represent historical data as the user-item interaction matrix $\mathbf{Y} \in \mathbb{R}^{m\times n}$,
where $\mathbf{Y}_{ij} =1$ indicates an observed interaction between user $i$ and item $j$, \eg a user reads a book or a user clicks a news; otherwise $\mathbf{Y}_{ij}=0$. 
In addition, we introduce knowledge graph $\mathcal{G}=\{(h,r,t)|h,t\in\xi, r\in\mathcal{R}\}$ as side information in the recommendation process, in which each triple $(h,r,t)$ indicates that there is a relationship $r$ from head entity $h$ to tail entity $t$, $\xi$ and $\mathcal{R}$ are the set of entities and relations in KG.

\noindent \textbf{Input:} User-item interaction matrix $\mathbf{Y}$ and knowledge graph $\mathcal{G}$.

\noindent \textbf{Output:} A novel duet representation learning framework for the recommendation task to correctly predict the probability $Pr(u,v)$ that user $u$ would adopt item $v$.

\section{Methodology}

\begin{figure*}[!t]
  {\centering\includegraphics[width=0.8\linewidth]{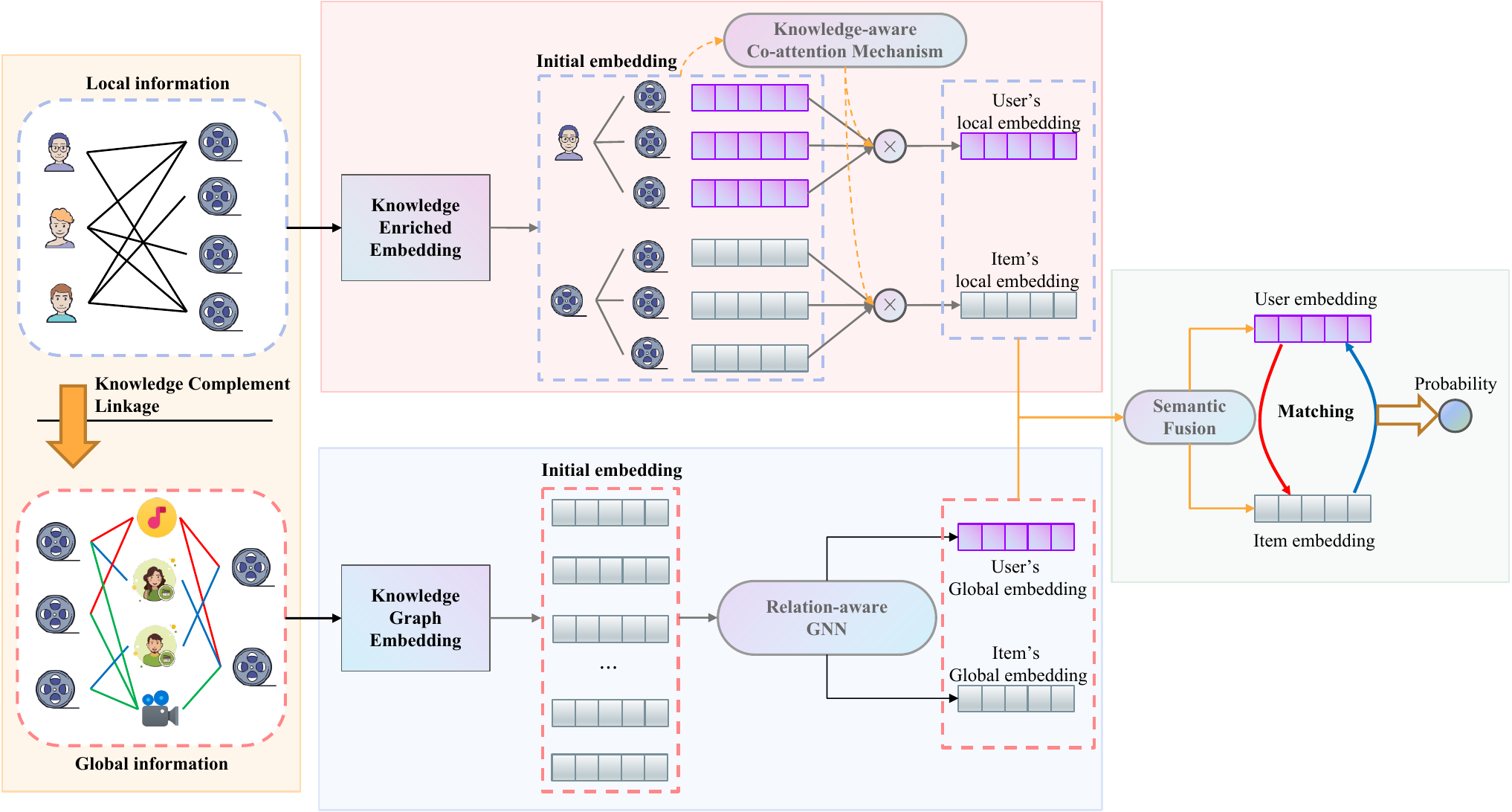}
    \caption{Illustration of the knowledge-aware duet model \sysname, which compromises four components: 
    i) Knowledge Complement Linkage (orange background, left), which maps items to external entities to capture rich semantic information in KG. 
    ii) Local Model (red background, middle), which learns local representations of users and items from the local information. 
    iii) Global Model (blue background, middle), which learns global representations of users and items based on the global information. 
    iv) Prediction (green background, right), which alleviates the semantic gap between the local and global information by a gated network and calculates the final predicted probability.}
  \label{fig:model-architecture}
  %\vspace{-0.2cm}
  }
\end{figure*}

In this section,  we will provide  details of the proposed knowledge-aware duet model \sysname, whose framework is illustrated in \figref{fig:model-architecture}.
The proposed model \sysname  comprises four key components for thoroughly learning the inherent characteristic of the local information and the global information.
First, knowledge complement linkage module is used for mapping items to external entities in KG to explore rich semantic information; 
Then, local model based on a knowledge-aware co-attention mechanism and global model based on a relation-aware GNN are designed to learn the representations of the users and items from the local and global information, respectively;
Finally, semantic fusion based on a gating network is developed to alleviate the semantic gap between both kinds of information, and the outputs are taken to compute the final predicted value in turn.
The remaining part of this section proceeds in terms of the four components mentioned above.

\begin{figure}[!t]
  \centering
  \includegraphics[width=0.4\textwidth]{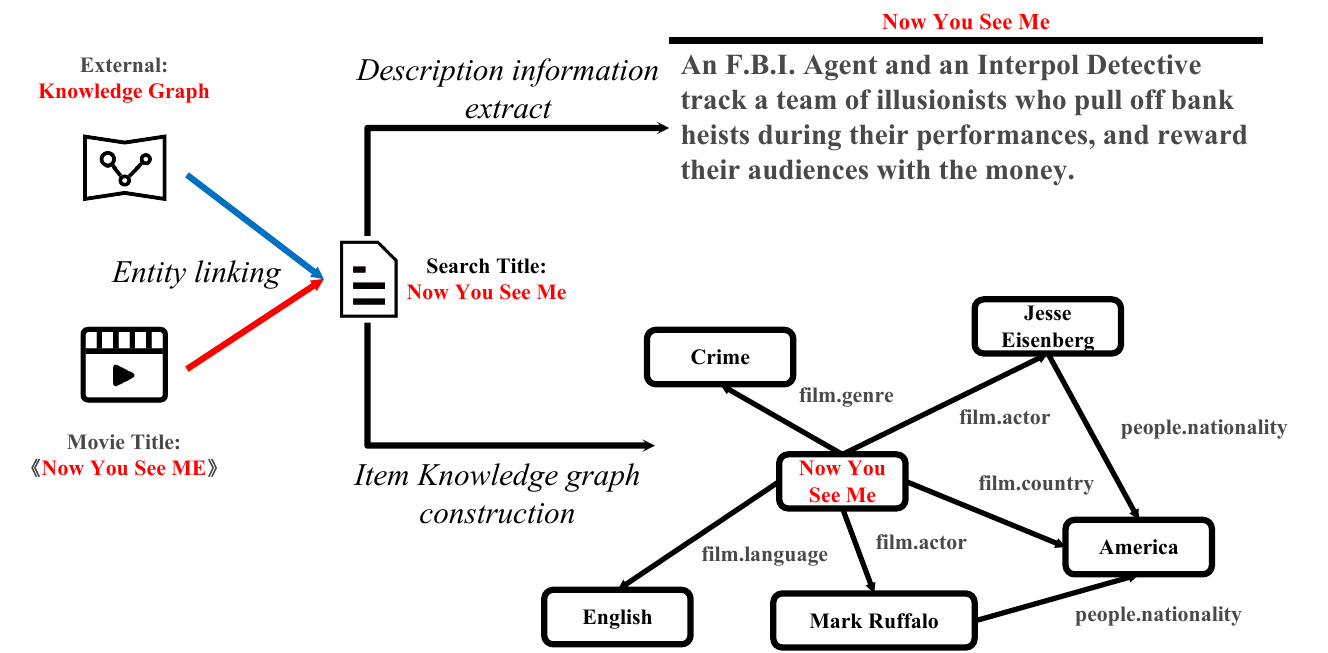}
  \caption{The process of knowledge complement linkage for each item, including entity linking, disambiguating entities, extracting items' description information and construct items' knowledge graph from external KGs.}
  \label{fig:Knowledge_complement}
  %\vspace{-0.1 in}
\end{figure}

\subsection{Knowledge Complement Linkage}
\label{sec:linkage}
The goal of knowledge complement linkage is to link each item with its corresponding entity in KG.
During this process, we first utilize entity linking \cite{milne2008learning, sil2013re} technology to retrieve related entities with items' titles as queries.
Occasionally, an item may have more than one entity returned during the linkage procedure. To address this problem, we further incorporate other items' attribute information to identify the accurate linkage entity (e.g., \textit{IMDB ID} and \textit{writer} \textit{name} are used for the movie and the book, respectively).
Based on the linked result, we can further extract the textual description of each entity and its centered subgraph as side information to enhance the recommendation process. 
For instance, as shown in \figref{fig:Knowledge_complement}, we take the title \textit{Now You See Me} as a query to retrieve the linked entity \textit{Now You See Me} in the external KG of a movie and further extract the textual description and subgraph based on its linked entity.

\subsection{Local Representation Learning Model}
The local model is proposed to explore the inner correlations between items and users based on local information.
As shown in \figref{fig:model-architecture}, it mainly includes three components: 
i) the Knowledge Enriched representation (KEE), which computes the initial representations for items;
ii) the encoding of users and items, which constructs the collaborative neighbor set and calculates the initial representation matrix; 
iii) the Knowledge-aware Co-attention Mechanism (KCM), which selects the most informative local neighbors for each user and item respectively. 

\begin{figure}[!t]
  \centering
  \includegraphics[width=0.4\textwidth]{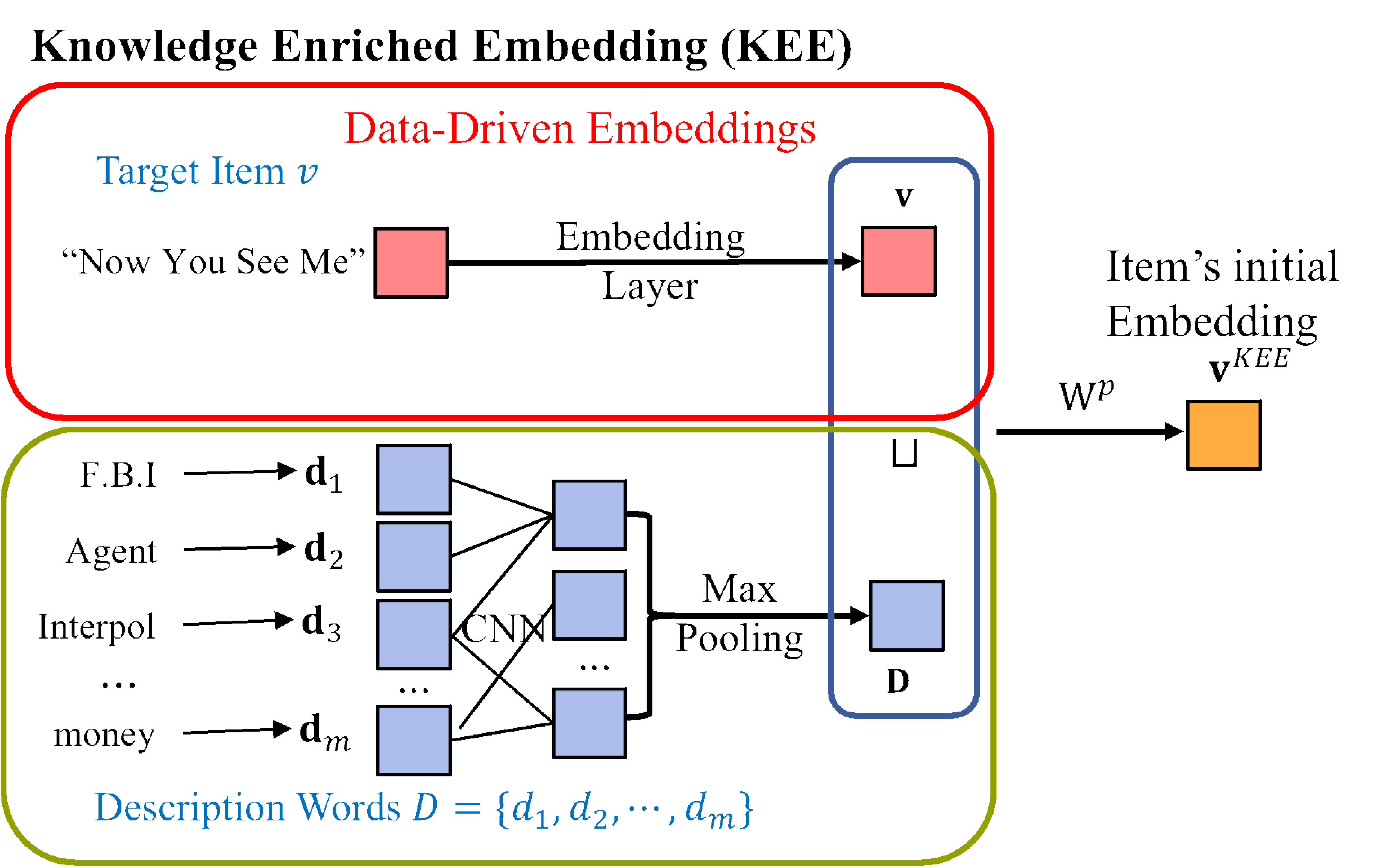}
  \caption{Illustration of Knowledge Enriched representation (KEE), which generates the enriched representation of items with their textual descriptions. }
  \label{fig:KG}
%\vspace{-0.1 in}
\end{figure}

\begin{figure*}
  \begin{minipage}[t]{0.48\textwidth}
    \centering
    \includegraphics[scale=0.3]{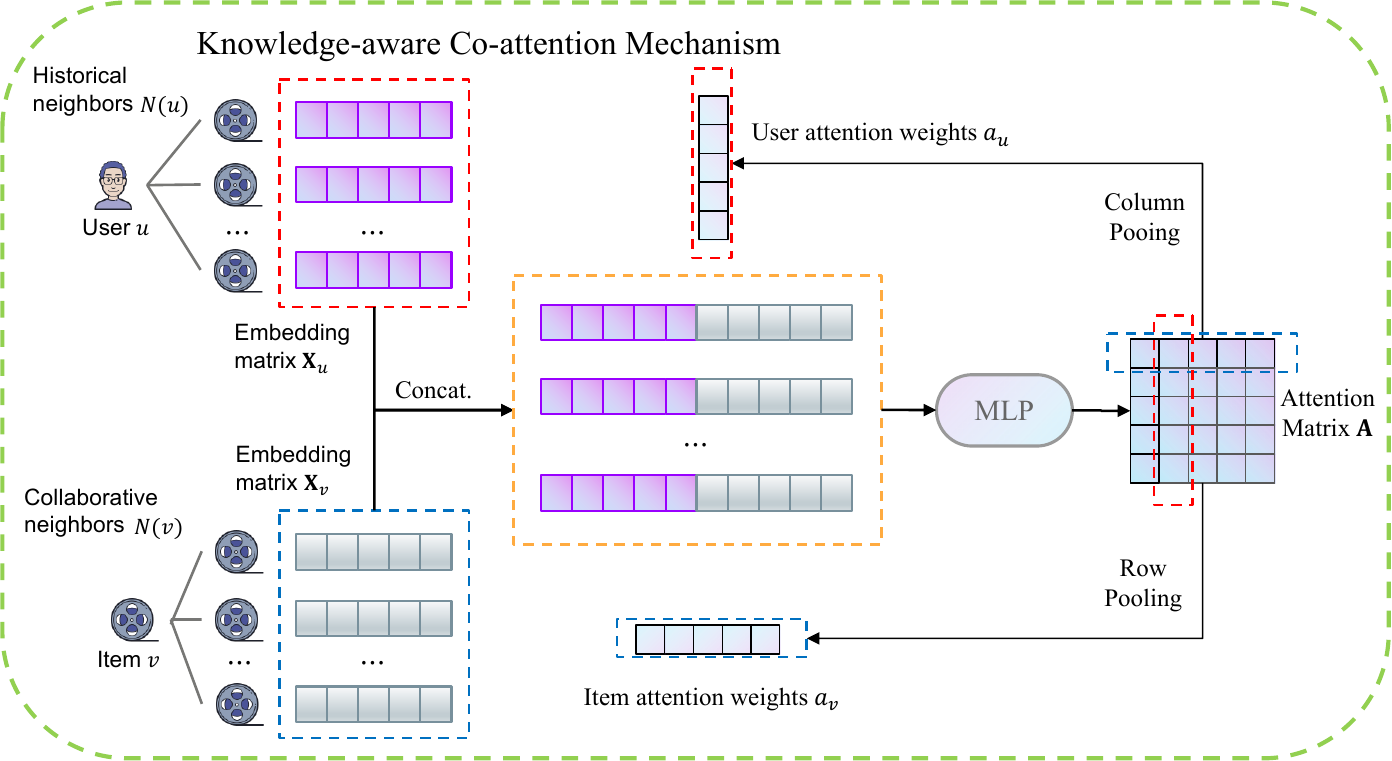}
    \caption{Illustration of  Knowledge-aware Co-attention Mechanism (KCM), which computes the corresponding local representation according to the historical and collaborative neighbors of users and items.}
    \label{fig:local_moel}
    \end{minipage}
    \hspace{0.1 in}
  \begin{minipage}[t]{0.48\textwidth}
    \centering
    \includegraphics[scale=0.27]{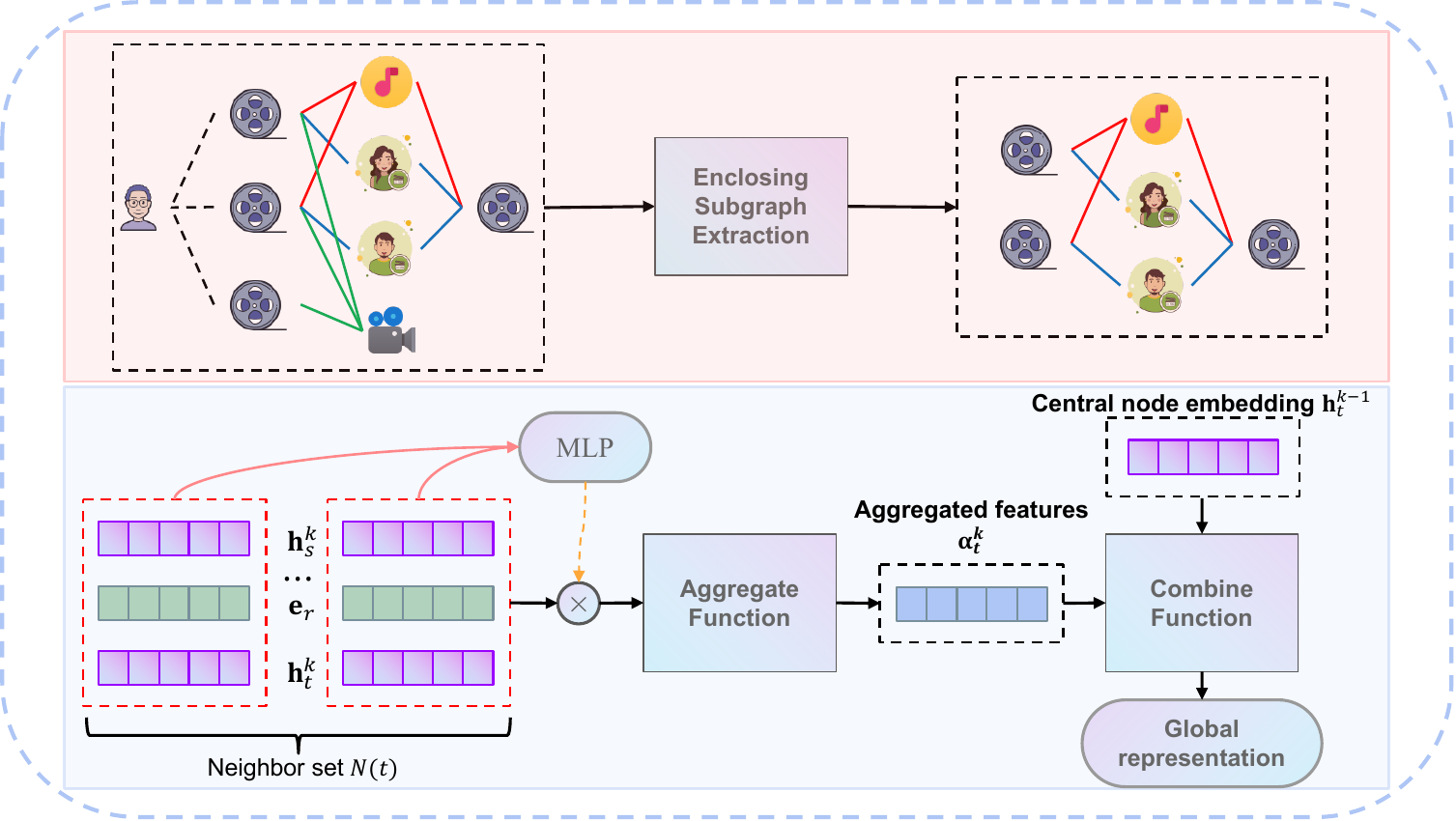}
    \caption{Illustration of Global Model, which contains three components the enclosing subgraph extraction, neural encoding of subgraph and relation-aware GNN.}
    \label{fig:relation-aware_GNN}
  \end{minipage}
  %\vspace{-0.2 in}
\end{figure*}

\fakeparagraph{Knowledge Enriched representation (KEE)}
Unlike the previous works \cite{hu2018local, zhang2016collaborative} using one-hot encoding or randomly generating representations, KEE enriches each item initial representation with its textual description and can be learned as part of local model.

As shown in \figref{fig:KG}, given the item $v$ and its textual description ${D} = \{d_{1}, d_{2}, \ldots, d_{m}\}$ composed of word sequences, 
KEE firstly uses a representation layer that maps the item into a representation with a lookup table \cite{xiong2017end}, then utilizes a CNN model to process the word sequence, which performs remarkably well for modeling sentence representation \cite{kim-2014-convolutional, xiong2018towards}.
Concretely, it embeds each word $d_{i}$ into $\mathbf{d}_{i}$ using representation layer, composes the word representations using CNN filters, and the maximum value of each dimension is obtained through max-pooling to generate the description representation $\mathbf{D}$.
And finally it is combined with the item representation $\mathbf{v}$ to learn the KEE representation $\mathbf{v}^{\text{KEE}}$ with a two-layer DNN $\mathcal{H}$.
The mathematical definition of KEE is shown as follows: 
\begin{displaymath}
    \begin{aligned}
        &v, d_{i} \stackrel{\mathbf{V}}{\longrightarrow} \mathbf{v}, \mathbf{d}_{i} & \textbf{representation Layer}\\
    	&F_q = \mathbf{W_i}\mathbf{d}_{p:p+h} &\textbf{CNN} \\
    	&\mathbf{D}_{i} = \max(F_1, \ldots, F_q, \ldots, F_{m-h}) &\textbf{Max-Pooling} \\
    	& \mathbf{D} = \{\mathbf{D}_1, \mathbf{D}_2, \ldots, \mathbf{D}_n\} & \textbf{Description representation} \\
    	&\mathbf{v}^{\text{KEE}} = \mathcal{H}(\mathbf{v} \sqcup \mathbf{D}) &\textbf{KEE representation}
    \end{aligned}
\end{displaymath}
where $\mathbf{V}$ is the parameters of the representation layer, $\mathbf{W_i}$ and $h$ are weights and size of the $i$-th filter, $n$ is the number of filters, and $\sqcup$ is the concatenation operator. 
Furthermore, with KEE, we can also obtain the initial representation for cold-start items without interaction information using their textual description.

\fakeparagraph{Encoding user and item} 
Rather than traditionally using an independent latent vector, we represent each user $u \in \mathcal{U}$ with a sequence of interacted items, which can be defined as $N(u) \in \mathbb{R}^{K_1\times 1}$, where $K_1$  represent the size of each user's neighbor set. 
Due to the unbalanced neighbor distribution between users and items, we utilize a neighbor selecting approach to select the top $K_1$ neighbors as users' neighbor collection, which is based on mutual information \cite{zhou2020improving} for ranking direct neighbors. 
The benefit of this approach is that it can reduce the redundancy in the neighbor set and keep the maximum retention of all neighbor features.

Analogously, items that have interacted with the same users in history can be considered similar to each other, which could be defined as collaborative neighbors.
Rather than treat items independently like most existing methods, we represent each item $v \in \mathcal{V}$ with its collaborative neighbor set $N(v) \in \mathbb{R}^{K_2\times 1}$, where $K_2$ represent the size of each item's neighbor set. 
During this process, the similarity or relevance between two items is calculated by the common users who have interacted with them \cite{wu2019dual}. 
Specifically, for any item pair $v_{i}$ and $v_{j}$, we define their similarity coefficient $s_{ij}$ as the proportion of users who interact with both items.
These coefficients induce an equivalence relation over items, \ie item $v_{i}$ is related to item $v_{j}$ if $s_{ij}>\tau$  with $\tau$ a fixed threshold.

For encoding collaborative signals in local information, unlike most methods considering one-side influence from users or items,  we respectively represented each user and item with its historical and collaborative neighbors extracted from local information.
Concretely, for one user-item pair $(u, v)$, we can represent them as $\mathbf{U} \in \mathbb{R}^{K \times 1}$ and $\mathbf{V} \in \mathbb{R}^{K \times 1}$, where $K$ is the size of neighbor set.
Following KEE, we transform each neighbor into a low-dimensional dense vector.
Therefore, we encode local neighbors of user $u$ and item $v$ into $\mathbf{X_u} \in \mathbb{R}^{K \times d}$ and $\mathbf{X_v} \in \mathbb{R}^{K \times d}$, where $d$ is the dimension for representation.

\fakeparagraph{Knowledge-aware Co-attention Mechanism (KCM)}
To select the most informative local neighbors for each user and item respectively and generate more meaningful representations of users and items, we propose a knowledge-aware co-attention module, which is shown in \figref{fig:local_moel}.
Given the local neighbors representation matrix of a user $\mathbf{X_u} \in \mathbb{R}^{K \times d}$ and an item $\mathbf{X_v} \in \mathbb{R}^{K \times d}$, we design an attention network with multiple layers to calculate an attention matrix $\mathbf{A} \in \mathbb{R}^{K \times K}$ as
\begin{equation}
    \mathbf{A}_{ij} = \mathbf{Attention}(\mathbf{X}^{i}_{u}, \mathbf{X}^{j}_{v}),
\end{equation}
where $\mathbf{X}^{i}_{u}$ is the $i$-th neighbor of user $u_{i}$, $\mathbf{X}^{j}_{v}$ is the $j$-th neighbor of item $v_{i}$, and $\mathbf{A}_{ij}$ is the calculated correlation value between them.
$\mathbf{A}$ contains the inner correlations among different collaborative neighbors.  
Furthermore, we respectively calculate the attention vectors of user or item by performing the mean-pooling operation along column or row on the attention matrix $\mathbf{A}$, which comprehensively considers the influence of different neighbors and can be defined as follows,
\begin{equation}
\begin{aligned}
    &\bm{a}_{u} = \textbf{Mean-Pooling}(\{A_{i\bm{\cdot}}\}^{K}_{i=1}),\\
    &\bm{a}_{v} = \textbf{Mean-Pooling}(\{A_{\bm{\cdot}j}\}^{K}_{j=1}),
\end{aligned}
\end{equation}
where $\bm{a}^{u} \in \mathbb{R}^{K \times 1}$ and $\bm{a}^{v} \in \mathbb{R}^{1 \times K}$ are the importance vectors for user $u$ and item $v$. 
After that, we take the normalization values of importance vectors as weights to calculate the local representations by attentively aggregating neighbor representation.
\begin{equation}
\begin{split}
    &\mathbf{u}^{\text{Local}}={\bm{a}'_{u}}^\mathrm{T} \mathbf{X}_{u},\ \bm{a}'_{u} = \sigma(\bm{a}_{u}),\\
    &\mathbf{v}^{\text{Local}}=\bm{a}'_{v} \mathbf{X}_{v},\ \bm{a}'_{v} = \sigma(\bm{a}_{v}),
\end{split}
\end{equation}
where $\sigma$ is the softmax function, $\mathbf{u}^{\text{Local}}$ and $\mathbf{v}^{\text{Local}}$ are the local representations of user $u$ and item $v$, respectively.
\subsection{Global Representation Learning Model}
Unlike most previous methods, which ignore the different impacts from different relation paths and may bring negative noises during conducting the information propagation process on the whole KG, the proposed global model consists of three components: 
i) Enclosing subgraph extraction, which extracts the enclosing subgraph for the given user-item pair from knowledge graph;
ii) Neural encoding of  subgraph, which learns a low-dimensional representation vector for each entity and relation that preserves the structural information of the graph;
iii) Relation-aware GNN, which reveals the different importance of relation-paths during information propagation on the subgraph.

\fakeparagraph{Enclosing subgraph extraction} 
Rather than existing methods capturing the knowledge associations on the whole graph, in this work, we only take the neighborhood of particular triplet in KG into consideration to eliminate noise during propagation.
To do so, we define \textit{enclosing subgraph} as the graph induced by all the entities that occur on a path between the two target entities.
For instance, ($e_0 \xrightarrow{produced \ by} e_1 \xrightarrow{produce} e_2 $) is a path included in enclosing subgraph around ($e_0, e_2$).

Hence, for extracting the enclosing subgraph of the given user-item pair $(u, v)$, we first represent user $u$ with its interacted items set $N(u)$.
Secondly, we construct the entity set $E(u)$ for user $u$ and target entity $e$ for item $v$ by mapping items into their corresponding entities in KG.
Thirdly, for each entity $e_i \in E(u)$, we compute the enclosing subgraph for $e_i$ and $e$ by taking the intersection of $\mathcal{N}_{k}(e_i)$ and $\mathcal{N}_{k}(e)$, which are set of $k$-hop neighbors, and further pruning the entities that are isolated from either entity $e_i$ or $e$.
Finally, as shown in \figref{fig:relation-aware_GNN}, we construct the enclosing graph for $(u, v)$ by merging all the enclosing subgraphs between every entity in $E(u)$ and target entity $e$.

\fakeparagraph{Neural encoding of subgraph} 
Knowledge graph representation (KGE) is an effective way to learn dense low-dimensional vector representations for entities and relations, which preserves the structural information of the graph.
Recently, translation-based KGE methods have received great attention due to their concise models and superior performance \cite{wang2018dkn}.
Therefore, in this paper, we employ a widely used translation-based method TransR \cite{lin2015learning}, which introduces a projection matrix for each relation to map entity representations to their corresponding relation space. 
Concretely, for each triple $(h, r, t)$ in the graph, we define that $\mathbf{e}_h, \mathbf{e}_t \in \mathbb{R}^{d \times 1}$, $\mathbf{e}_r \in \mathbb{R}^{k \times 1}$ and $\mathbf{M}_r \in \mathbb{R}^{k \times d}$ as the representations of entities $h$, $t$, relation $r$ and the projection matrix for relation $r$ respectively. 
The representation is learned by optimizing the translation principle $\mathbf{e}^{r}_{h} + \mathbf{e}_r \approx \mathbf{e}^{r}_{t}$ under the constraints of $\mathbf{e}^{r}_{h} = \mathbf{e}_h\mathbf{M}_r$ and $\mathbf{e}^{r}_{t}=\mathbf{e}_t\mathbf{M}_r$, which is the mapping representations of $\mathbf{e}_h$ and $\mathbf{e}_t$ in the relation $r$'s space. 
Then, the plausibility score for the triplet $(h, r, t)$ is defined as 
\begin{equation}
    g_{r}(h,t)=||\mathbf{e}^{r}_{h}+\mathbf{e}_{r} - \mathbf{e}^{r}_{t}||^{2}_{2}.
\end{equation}
The lower the plausibility score $g_{r}(h, t)$, the more valid the triplet is.
To encourage the discrimination between true and false triplets, we use the following margin-based ranking loss for training,
\begin{equation}
    \mathcal{L}_{KG} = \sum_{(h,h',r,t,t')\in \Gamma}{\max(0, g_{r}(h, t)+ \gamma - g_{r}(h', t'))},
\end{equation}
where $\gamma$ is the margin, and $\Gamma=\{(h,r,t,t')|(h,r,t)\in \mathcal{G}, (h', r, t') \notin \mathcal{G}\}$ is the training set for KGE. 

Following this way, we can initialize the representations of entities and relations on the granularity of triplets by exploiting the structural information in the extracted enclosing subgraph.

\fakeparagraph{Relation-aware GNN for Representation Learning}
The goal of global model is to compute the global representations of the user-item pair $(u, v)$ based on the enclosing subgraph extracted from global information.
We adopt the general message-passing scheme described in \cite{xu2018powerful}, where a node representation is iteratively updated by combining it with aggregation of its neighbors' representation.
Concretely, the $k$-th layer of GNN is given by
\begin{displaymath}
    \begin{aligned}
        &\mathbf{a}_{i}^{k} = \mathbf{AGGREGATE}^{k}({\mathbf{h}_{s}^{k-1}:s\in \mathcal{N}(t)}, \mathbf{h}_{t}^{k-1}),\\
    	&\mathbf{h}_{t}^{k}=\mathbf{COMBINE}^{k}(\mathbf{h}_{t}^{k-1}, \mathbf{a}_{t}^{k}),
    \end{aligned}
\end{displaymath}
where $\mathcal{N}(t)$ is the set of immediate neighbors of node $t$, $a_{t}^{k}$ denotes the aggregated message from the neighbors, and $h_{t}^{k}$ denotes the latent representation of node $t$ in the $k$-th layer.
During the message-passing process, the initial latent representation $h_{i}^{0}$ of any node $i$ is initialized by neural encoding of subgraph with KGE in previous components. 
Furthermore, this framework gives the flexibility to plug in different $\mathbf{AGGREGATE}$ and $\mathbf{COMBINE}$ functions resulting in various GNN architectures \cite{teru2020inductive}.

Inspired by the fact that message passed from different relation-path in graph may have different influence and the multi-relational R-GCN \cite{schlichtkrull2018modeling}, we design a relation-aware GNN, as shown in \figref{fig:relation-aware_GNN}, whose \textbf{AGGREGATE} function is defined as
\begin{equation}
    \mathbf{a}_{t}^{k+1} = \sum_{r=1}^{\mathcal{R}}\sum_{s\in \mathcal{N}_{r}(t)}{\omega_{rst}^{k+1}}\mathbf{W}^{k+1}_{r}\mathbf{h}^{k}_{s},
\end{equation}
where $\mathcal{R}$ is the total number of relations present in graph, $\mathcal{N}_{r}(t)$ is the neighbors of node $t$ connected with relation $r$, $\mathbf{W}^{k+1}_{r}$ is the transformation matrix for relation $r$ used to propagate messages in the $k$-th layer, and $\omega_{rst}^{k+1}$ is the relation-path attention weight at the $k$-th layer corresponding to the triplet $(s, r, t)$, which is calculated via a two-layer MLP as follow:
\begin{equation}
\begin{aligned}
    &\mathbf{c} = \mathbf{ReLU}(\mathbf{W}_{1}[\mathbf{h}_{s}^{k}\oplus \mathbf{h}_{t}^{k}\oplus \mathbf{e}_{r}]+\mathbf{b}_{1}), \\
    &\omega_{rst}^{k+1} = \sigma(\mathbf{W}_{2}\mathbf{c}+\mathbf{b}_{2}),
\end{aligned}
\end{equation}
where $\mathbf{h}_{s}^{k}$ and $\mathbf{h}_{t}^{k}$ are the latent representation of head and tail node of a triplet at $k$-th layer of the GNN, $\mathbf{e}_{r}$ is the learned representation of relation $r$ in graph, and $\sigma$ is a sigmoid function which regulates the information aggregated from each neighbor.
In practice, to avoid overfitting on rare relations caused by the rapid growth in the amount of parameters with the number of relations in the graph, we adopt the basis sharing mechanism among the relation-specific transformation matrices $\mathbf{W}^{k+1}_{r}$ of each layer in GNN and also implement a form of triplet dropout, where triplets are randomly dropped from the graph while aggregating information from the neighborhood.
Furthermore, given the aggregated information $\mathbf{a}_{t}^{k+1}$ of node $t$, we implement the \textbf{COMBINE} function with a self-connection of a special relation type to compute the updated representation for it, derived from \cite{schlichtkrull2018modeling}. 
It is given by
\begin{equation}
    \mathbf{h}_{t}^{k+1} = \mathbf{LeakyReLU}(\mathbf{W}_{3}\mathbf{h}^{k}_{t}+\mathbf{a}^{k+1}_{t}).
\end{equation}
Following the relation-aware GNN architecture as described above, we obtain the global representations of items after $L$ layers of message passing on the corresponding subgraph.
Then, we compute the global representation of user with its interacted items as follows,
\begin{equation}
\begin{aligned}
    &\mathbf{v}^{\text{Global}} = \mathbf{h}_{v}^{L}, \\
    &\mathbf{u}^{\text{Global}} = \frac{1}{|\mathcal{N}(u)|}\sum_{v_i \in \mathcal{N}(u)}{\mathbf{h}_{v_i}^{L}},
\end{aligned}
\end{equation}
where $|\mathcal{N}(u)|$ is the size of item set, $\mathbf{v}^{\text{Global}}$ and $\mathbf{u}^{\text{Global}}$ are global representations of item $v$ and user $u$, respectively.

\subsection{Prediction with Semantic Fusion}
Note that local and  global information could jointly indicate user preference on item, but for distinct user-item pairs, the importance of the two-fold effects could be different.
Therefore, inspired by \cite{zhou2020improving}, we design a gating network for semantic fusion to dynamically allocate weights to the four semantic features ($\mathbf{v}^{\text{Local}}$, $\mathbf{u}^{\text{Local}}$, $\mathbf{v}^{\text{Global}}$, $\mathbf{u}^{\text{Global}}$) according to specific user-item pair.

\fakeparagraph{Semantic fusion with gating network} For a user-item pair $(u, v)$, given the local and global representations of them from the above two sub-models, we could derive the final representations with the gating network as follows, 
\begin{equation}
\begin{split}
    \mathbf{u}^{\text{Final}} &= \alpha\cdot \mathbf{u}^{\text{Global}} + (1-\alpha )\cdot \mathbf{u}^{\text{Local}}, \\
    \alpha &= \sigma(\mathbf{W}_{\text{gate}}[\mathbf{u}^{\text{Global}} \sqcup \mathbf{u}^{\text{Local}}]),
    \end{split}
\end{equation}
where $\mathbf{u}^{\text{Final}}$ is the final representation of user $u$, $\alpha$ denote the weight for global features and $\mathbf{W}_{\text{gate}}$ is a learnable transformation  matrix of the linear unit. 
The final representation $\mathbf{v}^{\text{Final}}$ of item $v$ could also be computed in similar way.
Taking the final representation as input, we can compute the probability that user $u$ will adopt item $v$,
\begin{equation}
    Pr(u, v) = nn(\mathbf{u}^{\text{Final}}, \mathbf{v}^{\text{Final}}),
\end{equation}
where $nn(\cdot)$ can be a fully-connected network with a sigmoid activation function.

\fakeparagraph{Loss function} To optimize the recommendation model, we adopt the loss function of BPR loss \cite{rendle2012bpr}, which assumes that the observed interactions that indicate more user preferences should be assigned higher predictions values than unobserved ones:
\begin{equation}
    \begin{split}
    Loss = -\sum_{(u, v_i, v_j)\in \mathcal{O}}{ -\ln\sigma (Pr(u, v_i) - Pr(u, v_j)) + \lambda ||\Theta||^{2}_{2}},
    \end{split}
\end{equation}
where $\mathcal{O}=\{(u, v_i, v_j)|(u,v_{i})\in \mathcal{I}^{+}, (u, v_{j}) \in \mathcal{I}^{-}\}$ denotes the training set,  $\mathcal{I}^{+}$ indicates the positive interactions between user $u$ and items while $\mathcal{I}^{-}$ is the sampled negative interaction set, $\sigma(\cdot)$ is the sigmoid function, and $\Theta$ is the model parameter set. $L_2$ regularization parameterized by $\lambda$ on $\Theta$ is conducted to prevent overfitting. 
To optimize the loss function, we adopt the mini-batch Adam in our implementation for its ability to adaptively control the learning rate.

\subsection{Time Complexity of \sysname}
For the local model, the computational complexity is $O(n(m-h)hd+d(n+d))$ for KEE operation, and $O(2K(d+1))$ for KCM operation, where $n$ and $m$ are the number of filters and fixed length of description content, $h$ is the filter size in KEE, $K$ and $d$ are the fixed neighbor size and length of embeddings, respectively, so the computational complexity in this part is $O(n(m-h)hd+d(n+d)+2K(d+1))$.
For the global model, the computational complexity for enclosing subgraph extraction is $O((K+1)(|\xi|+|E|))$ while utilizing Breadth-First Search algorithm to prune the isolated entities, and $O(|\mathcal{R}|dk)$ for representation learning.
where $|E|$, $|\xi|$ and $|\mathcal{R}|$ are the size of edges, nodes and relations in the graph, respectively, $d$ is the size of node/relation embeddings.
It can be seen that the computational cost of the global model depends on the size of the graph, which can be greatly reduced by extracting enclosing subgraphs.
In the semantic fusion part, the computation complexity of gating network is $O(d)$, and the computational complexity of the last fully-connected layer is $O(d^2)$.

\section{EXPERIMENTS}
To comprehensively evaluate the proposed model \sysname, we conduct experiments to answer the following research questions:

\fakeparagraph{RQ1} How does \sysname perform compared with state-of-the-art models for recommendations, especially the KG-based recommendation models?

\fakeparagraph{RQ2} How do hyper-parameters and the key components in \sysname impact the recommendation performance?

\fakeparagraph{RQ3} Could \sysname provide some reasonable explanations about user preferences benefiting from the knowledge graph and attention mechanism?

\subsection{Dataset Description}
To evaluate the effectiveness of \sysname, we apply our model to two public benchmark datasets Movielens and Last-FM. 
The statistics of these two datasets are shown in \tabref{table:1}. 
The basic descriptions about them are summarized as follows:
\begin{itemize}
	\item \textbf{MovieLens-1M}\footnote{https://grouplens.org/datasets/movielens/} contains approximately 1 million explicit ratings (ranging from 1 to 5) on the MovieLens website. 
	We extract the 10-core data to ensure data quality. 
	\item \textbf{Last.FM}\footnote{https://grouplens.org/datasets/hetrec-2011/} contains musician listening information from a set of 2 thousand users from Last.FM online music system. 
	Similarly, we use the 10-core setting to ensure that each user and item pair has at least ten interactions.
\end{itemize}

\begin{table}[!t]
   \setlength{\abovecaptionskip}{0.1cm}
   \setlength{\belowcaptionskip}{0cm}
    \caption{Basic statistics of the datasets.}
    \centering
	\begin{tabular}{p{1.3cm}p{1.7cm}p{2.2cm}<{\centering}p{1.8cm}<{\centering}}
  	\bottomrule
		 \quad &\quad & MovieLens-1M & Last.FM\\
	\end{tabular}
	\begin{tabular}{p{1.3cm}p{1.7cm}p{2.2cm}<{\centering}p{1.8cm}<{\centering}}
    	\hline
		\multirow{3}{*}{\makecell[c]{User-Item \\Feedback}} 
		&\#Users & 6,040 & 1,851 \\
		&\#Items & 3,389 & 2,315 \\
		&\#Interactions & 997,024 & 59,781\\
%		\midrule
	\end{tabular}
	\begin{tabular}{p{1.3cm}p{1.7cm}p{2.2cm}<{\centering}p{1.8cm}<{\centering}}
    	\hline
		\multirow{3}{*}{\makecell[c]{ Knowledge\\Graph}} &  \#Entities & 392,966 & 10,367\\
		&\#Relations & 49 & 63\\
		&\#Triplets & 2,112,838 & 245,043 \\
%		\midrule
		\bottomrule
	\end{tabular}
	\label{table:1}
	%\vspace{-0.2 in}
\end{table}

In order to be consistent with the implicit feedback setting, we transform them into implicit feedback where each user-item pair is marked with 1 indicating that the user has rated the item positively.
The threshold of positive for MovieLens-1M is 4, while no threshold is set for Last.FM due to their sparsity.

Besides the user-item interactions, we need to construct a knowledge graph for each dataset. 
Concretely, we follow the way in Knowledge Complement Linkage (\secref{sec:linkage}) to map items to Freebase entities\footnote{https://developers.google.com/freebase/} via title matching.
For those items that failed to link, we simply discard them.
Furthermore, for identified entities, we consider
the triplets that are immediate neighbors of the entities aligned with items, no matter which role (\ie subject or object) it serves as.
To ensure the KG quality, we then filter out infrequent entities expected for entities aligned with items (\ie lower than 10 in both datasets) and retain the relations appearing in at least 50 triplets. 
The basic statistics of the extracted knowledge graph information for the two datasets are also summarized in \tabref{table:1}.

For each dataset, we randomly select $80\%$ of interaction history of each user to constitute the training set, and treat the remaining as the test set. 
From the training set, we randomly select $20\%$ of interactions as validation set to tune hyper-parameters.
For each observed user-item interaction, we treat it as a positive instance, and then conduct the negative sampling strategy to pair it with one negative item that the user did not rate before.

\subsection{Experiment Setup}

\fakeparagraph{Evaluation Metrics}
For each user in the test set, we randomly sample 100 items that the user has not interacted with as the negative items, considering the computational efficiency. 
Then each method outputs the user’s preference scores over all the items in test environment. 
To evaluate the effectiveness of top-K recommendation and preference ranking, we adopt two widely-used evaluation protocols: recall@K and ndcg@K \cite{chang2021sequential}.
By default, we set K = 20. 
We report the average metrics for all users in the test set.

\fakeparagraph{Baselines}
To evaluate the effectiveness of \sysname, we compare the proposed model with CF-based (FM and NFM), regularization-based (CKE and CFKG), and GNN-based (KGAT, MVIN and CKAN):
\begin{itemize}
	\item \textbf{FM} \cite{rendle2011fast}: A basic factorization method for modeling the second-order feature interactions between inputs. In our evaluations, we treat ID of a user, an item, and the related KG knowledge as input features.
	\item \textbf{NFM} \cite{he2017neural}: The method is a state-of-the-art factorization model, which subsumes FM under neural network. 
	Specially, we enrich the representation of an item with the embeddings of connected entities in KG and employ one hidden layer on input features as suggested in \cite{he2017neural}.
	\item \textbf{CKE} \cite{zhang2016collaborative}: It combines CF with various information, including structural, textual, and visual knowledge in a unified framework for the recommendation. 
	We implement CKE as CF plus structural knowledge in this paper.
	\item \textbf{CFKG} \cite{ai2018learning}: It applies TransE \cite{bordes2013translating} on the unified graph including users, items, entities, and relations, transforming the recommendation task into the plausibility prediction of (user, Interact, item) triplets.
	\item \textbf{KGAT} \cite{wang2019kgat}: It employs a graph attention network on a unified graph, which includes the knowledge graph and user-item graph, to discriminate the importance of neighbors in graph.
	\item \textbf{MVIN} \cite{tai2020mvin}: It learns the representations of items from both the user view and the entity view. 
	MVIN gathers the knowledge in KG and the different interactions between entities to model user preference.
	\item \textbf{CKAN} \cite{wang2020ckan}: It encodes the collaborative signals that are latent in user-item interactions and combines them with KG in an end-to-end manner. 
	CKAN initialize the entity set of user and item with the collaborative signals.
\end{itemize}

\fakeparagraph{Implement Details}
All models are implemented based on PyTorch, in which the hyper-parameters are configured following popular choices or previous research.
In detail, we optimize all models with Adam \cite{kingma2014adam} optimizer, where the batch size is fixed at 512.
The default Xavier initializer \cite{glorot2010understanding} is used to initialize the model parameters. 
We apply a grid search for some common hyper-parameters in all models: 
the learning rate is tuned amongst \{0.05, 0.01, 0.005, 0.001\} with the decay rate of 0.9, the coefficient of $L_2$ normalization is searched in $\{10^{-5}, 10^{-4}, 10^{-3}, 10^{-2}\}$, and the dropout ratio is tuned in \{0.2, 0.3, 0.4, ..., 0.8\}.
For other hyper-parameters of baselines, the settings are the same as reported in their original papers or as default in their codes.
Then for the hyper-parameters specific to our proposed model, they are set as follows:
The size of neighbors is set at 40 for MovieLens-1M and 20 for Last.FM according to their distribution. 
For the local model, the dimension of word embeddings is 64 and is initialized with Word2vec \cite{mikolov2013efficient} using wiki corpus.
The length of the CNN with 64 filters used to encode description is set to 3, which refers to tri-gram. 
The fixed size of descriptions is set 30 for MovieLens-1M and 40 for Last.FM according to their distribution. 
For the global model, we choose TransR \cite{lin2015learning} to learn the pre-trained entity and relation embedding with the dimension of 128.
In consideration of computational efficiency, we limit the number of triples where the entity is located to 1,000 for Last.FM and 100 for MovieLens-1M and preserve 2-hop neighbors during constructing the enclosing subgraph.
The codes of this paper are available at \href{https://github.com/scwu1008/KADM}{https://github.com/scwu1008/KADM}.

\begin{table}[!t]
   \setlength{\abovecaptionskip}{0.1cm}
   \setlength{\belowcaptionskip}{0cm}
    \caption{Comparative results of MovieLens-1M and Last.FM. For Recall, NDCG, the larger value is better.}
    \centering
% 	\begin{tabular}{p{1.3cm}p{1.7cm}|p{2.2cm}<{\centering}|p{1.8cm}<{\centering}}
%     	\hline
% 		 \quad &\quad & MovieLens-1M & Last.FM\\
% 		\hline
% 	\end{tabular}
	\begin{tabular}{p{1.3cm}<{\centering}p{1.4cm}<{\centering}p{1.4cm}<{\centering}p{1.3cm}<{\centering}p{1.3cm}<{\centering}}
    	\bottomrule
		\multirow{2}{*}{Model} 
		&\multicolumn{2}{c}{Last.FM} & \multicolumn{2}{c}{MovieLens-1M} \\
 		\cline{2-3} \cline{4-5}
		& recall & ndcg & recall & ndcg \\
		\hline
%		\midrule
		%\bottomrule
	\end{tabular}
	
    \begin{tabular}{p{1.3cm}<{\centering}p{1.4cm}<{\centering}p{1.4cm}<{\centering}p{1.3cm}<{\centering}p{1.3cm}<{\centering}}
    	%\hline
		FM & 0.568 & 0.448 & 0.534 & 0.610 \\
% 		\hline
		NFM & 0.535 & 0.412 & 0.590 & 0.620 \\
		\hline
		CKE & 0.553 & 0.483 & 0.635 & 0.670 \\
% 		\hline
		CFKG & 0.577 & 0.484 & 0.621 & 0.672 \\
		\hline
		KGAT & 0.657 & 0.550 & 0.652 & 0.701 \\
% 		\hline
		MVIN & \underline{0.672} & \underline{0.583} & \underline{0.658} & \underline{0.713} \\
		CKAN & \underline{0.686} & \underline{0.590} & \underline{0.673} & \underline{0.721} \\
		\hline
		\textbf{KADM} & \textbf{0.736 }& \textbf{0.625 }& \textbf{0.694 }&\textbf{0.752} \\
		% \sysname & ($+0.050~0.201$) & ($+0.035~0.213$) & ($+0.021~0.160$) & ($+0.031~0.142$)
		%\hline
		\bottomrule
	\end{tabular}
	\label{table:2}
	%\vspace{-0.2 in}
\end{table}

\subsection{Performance Comparison (RQ1)}
The experiment  results for algorithm overall comparison are shown in \tabref{table:2}, we have some observations from it:
\begin{itemize}
    \item \textbf{Our proposed \sysname has the best performance in all metrics on both two datasets.}
    Overall, \sysname surpasses others significantly by around 0.05 and 0.04 on Last.FM dataset and by 0.02 and 0.03 on MovieLens-1M dataset in metrics of recall and ndcg at least, respectively.
    It indicates that \sysname has the significant power of explicitly encoding collaborative signals with the co-attention manner and capturing the rich knowledge associations contained in global information with the relation-aware attention mechanism. 
    From another aspect, the combination of collaborative signals in local information and knowledge associations in global information can obviously improve the  recommendation performance. 
    Meanwhile, KADM has better performance than CKAN on both datasets, which indicates that the dual mechanism is effective for modeling different kinds of information.
    \item \textbf{GNN-based models have better performances than other kinds of baselines, but are affected by introduced noise.}
    It illustrates that capturing the information propagation on KG with GNN can be effective to model user preference.
    However, when the graph becomes denser and larger, more noise would be introduced to the propagation process \cite{tai2020mvin}. 
    Therefore, directly computing on the enclosing subgraph between user and item in \sysname, which only contains the relevant entities and relation-paths in large-scale KG, can be more effective.
    \item \textbf{Most KG-based methods perform better than traditional CF-based methods on all datasets.}
    It demonstrates that the usage of KG is of great help for the recommendation. 
    Meanwhile, it is worth noting that the performances of the GNN-based models are better than regularized-based models, which indicates that modeling the first-order relationship might not fully utilize the structural information in KG. 
    %Instead, modeling high-order knowledge information is effectiveness.
    \item \textbf{In most situations, the model performance on movie data is better than music data.}
    One possible assumption is that there are more interactions between users and items in local information and links among nodes in global information, which provides sufficient information for learning the latent embeddings.
\end{itemize}

\begin{figure}[!t]
  \centering
  \includegraphics[width=0.47\textwidth]{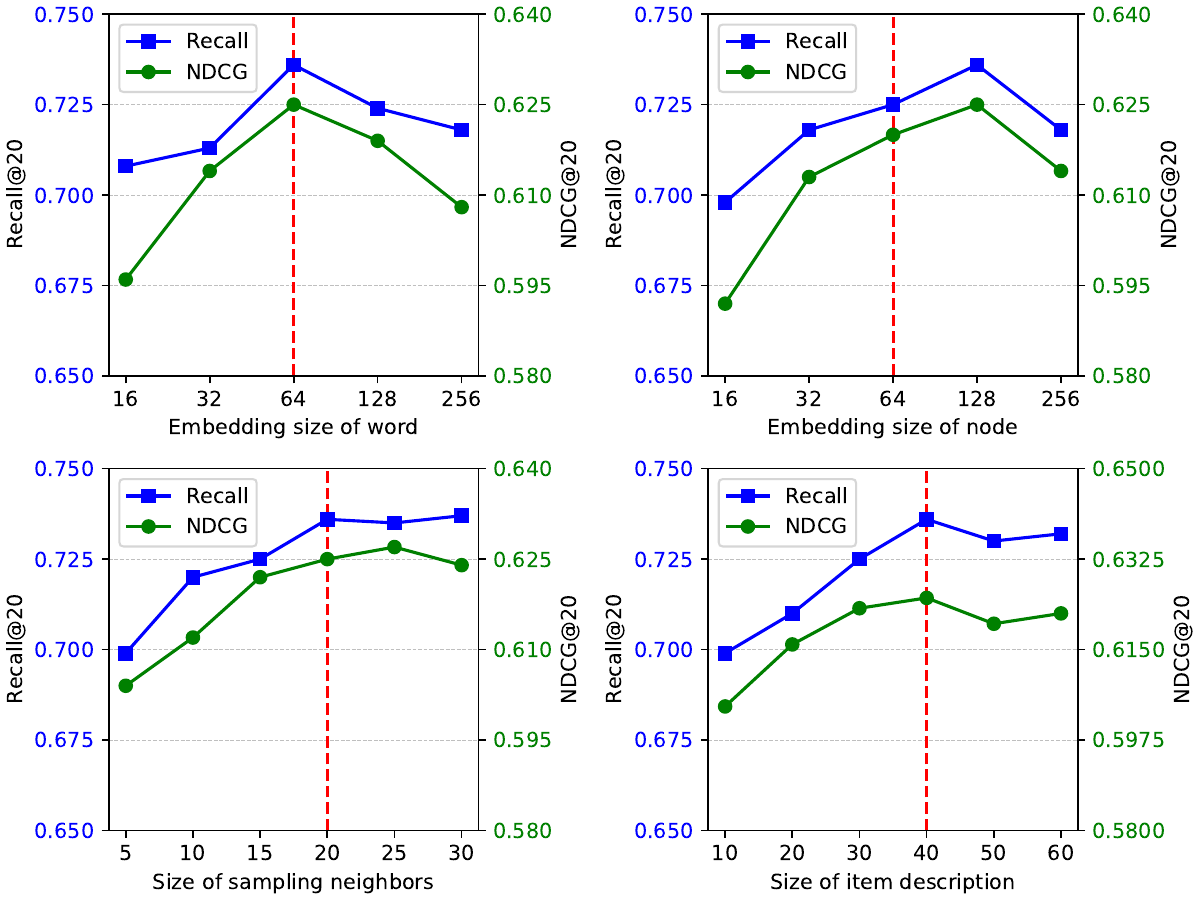}
  \caption{Evaluation of \sysname on Last.FM w.r.t different hyper-parameters. 
  }
  \label{fig:hyper-parameter}
  %\vspace{-0.1 in}
\end{figure}

\subsection{Study Of \sysname (RQ2)}
To study the performance variation for our model, we conduct some experiments on Last.FM with different hyper-parameter settings.

\fakeparagraph{Effect of dimension of embedding} 
In \sysname, we explore the impact of different dimensions on the model performance, including word embedding in  local model and node embedding in  global model. 
The results are shown in \figref{fig:hyper-parameter}, which enlightens us that a proper embedding dimension is needed. 
If it is too small, it would lack expressiveness; while if it is too large, it might cause a little overfitting, which leads to performance decline.

\fakeparagraph{Effect of sampling neighbor size}
We vary the size of sampling neighbor to investigate the impact of usage of the local and global information, which influences the initial neighbor set of user and item in \sysname. 
From \figref{fig:hyper-parameter}, we observe that the performance would be better with the sample size increasing, but the rate of increase is decreasing and the computational cost increases as well.
Therefore, we can select an appropriate sample size that can keep a good balance between model performance and complexity.

\fakeparagraph{Effect of item description size}
The change of description size may influence the learned initial embedding of items in local model. 
To investigate its impact, we experiment with different sizes of description $D$.
From \figref{fig:hyper-parameter}, it illustrates that the model performance first increases and then decreases as the length increases.
A reasonable explanation is that a too small $D$ lacks enough capacity to describe the item, while a too large $D$ is prone to be misled by noises.

% \fakeparagraph{Effect of depth of subgraph}

\begin{table}[!t]
   \setlength{\abovecaptionskip}{0.1cm}
   \setlength{\belowcaptionskip}{0cm}
    \caption{Effect of different network configurations.}
    \centering
	\begin{tabular}{p{1.7cm}<{\centering}p{1.3cm}<{\centering}p{1.3cm}<{\centering}p{1.2cm}<{\centering}p{1.2cm}<{\centering}}
    	\bottomrule
		\multirow{2}{*}{Model} 
		&\multicolumn{2}{c}{Last.FM} & \multicolumn{2}{c}{MovieLens-1M} \\
 		\cline{2-3} \cline{4-5}
		& recall & ndcg & recall & ndcg \\
	\end{tabular}
	
    \begin{tabular}{p{1.7cm}<{\centering}p{1.3cm}<{\centering}p{1.3cm}<{\centering}p{1.2cm}<{\centering}p{1.2cm}<{\centering}}
    
    	\hline
		\small \sysname-co & 0.704 & 0.613 & 0.669 & 0.719 \\
		\small \sysname-rel & 0.695 & 0.607 & 0.662 & 0.714 \\
		\small \sysname-local & 0.674 & 0.572 & 0.654 & 0.708 \\
		\small \sysname-global & 0.689 & 0.603 & 0.663 & 0.716 \\
		\small \sysname & 0.736 & 0.625 & 0.694 & 0.752 \\
		\bottomrule
	\end{tabular}
	\label{table:3}
	%\vspace{-0.2 in}
\end{table}

\fakeparagraph{Effect of different network configurations}
In order to verify the effectiveness of some components in our model, we conduct some ablation studies and the results are shown in \tabref{table:3}. 
There are three different network configurations used for ablation experiments: 
1) Co-attention mechanism, which can compute the inner correlation in local information. In \sysname-co, we fix the attention weight in the local model to $\frac{1}{K}$ , where $K$ is the size of the sampling neighborhood;
2) Relation-aware attention mechanism, which models the different influence of relation paths during the information propagation process; 
In \sysname-rel, we modify the $\mathbf{AGGREGATE}$ function in the global model to an averaging function, and take the mean of the immediate neighbors as the aggregated feature.
3) Combination of local and global models, which complement each other for modeling user preferences. 
We conduct experiments on the local model and the global model after removing gate network settings, separately.
As we can see, the results enlighten us that:
i) The attention mechanisms are both effective to help us filter useful features from local and global information, which is beneficial for modeling user preference;
ii) This supports our underlying hypothesis that modeling user preference with both local and global information could complement each other to achieve a better performance, and hence a combination of them is more appropriate.

\begin{figure}[!t]
  \centering
  \includegraphics[width=0.47\textwidth]{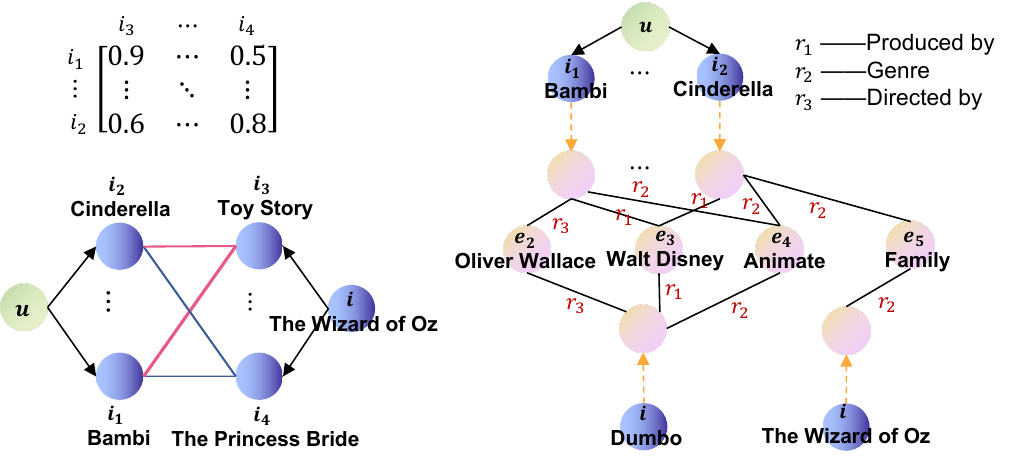}
  \caption{A real example from MovieLens-1M, including local information (left) and global information (right).}
  \label{fig:case_study}
  %\vspace{-0.2 in}
\end{figure}

\subsection{Case Study (RQ3)}
Benefiting from the attention mechanism and knowledge graph, we can reason on the attention matrix in local model and the high-order connectivity of enclosing subgraph in global model to infer the user preferences on the candidate item, offering some potential explanations.
Specially, we conduct a case study of one user-item pair and show the computed attention matrix and the extracted higher-order connectivity of enclosing subgraph between user $u$ and item $i$ in \figref{fig:case_study}. 
There are three key observations: 
a) For local information, \sysname computes the attention matrix from it, which is beneficial to infer the user preference. The attention weights between collaborative neighbors of user $u$ and item $i$ can be considered as  evidence of why the item meets the user preference. 
As we can see, most neighbors of item $i$ are similar to neighbors of user $u$, which denotes that the target item matches the user taste.
b) For global information, \sysname captures the high-order connectivity of enclosing subgraph extracted from it, which can also play an important role to infer user preference. 
The connected paths in the subgraph are useful for inferring user preference.
For instance, the connectivities ($Bambi \xrightarrow{r_3} Oliver Wallace  \xrightarrow{r_3} Dumbo $) and ($Bambi \xrightarrow{r_1} Walt Disney \xrightarrow{r_1} Dumbo $) indicate that the target item \textit{Dumbo} has the same director and publisher as the user's favorite movie \textit{Bambi} in the past. 
Hence, we can generate the explanation as \textit{Dumbo} is recommended since you have watched \textit{Bambi} directed by the same director \textit{Oliver Wallace} and produced by the same producer \textit{Walt Disney}.
c) For the whole model, the local model and the global model can complement each other. 
    Concretely, even if the target item itself has almost no related path connected with the user in global information, we can still recommend the item through its collaborative neighbor similarity with the user, and vice versa.
    For example, we would recommend \textit{The Wizard of Oz} even though the subgraph is sparse, because the reason that its collaborative neighbors are very similar to the user's.
\section{Conclusion and Future Work}
In this work, we study the task of combining the user-item interaction data and knowledge graph information for top-$N$ recommendation.
We propose a duet representation learning framework \sysname to unify the local and global information, in which a local model with a knowledge-aware co-attention mechanism is developed to learn the local representation of items and users by discovering the inner correlations from their collaborative neighbors, and a global model with relation-aware GNN is designed to learn the global representation of items and users by capturing the knowledge associations in the enclosing subgraph from knowledge graph.
Extensive experiments on two real-world datasets verify the effectiveness of \sysname.
Further evaluations on hyper-parameters and case studies also demonstrate the advantages of \sysname.

As for future work, we would make focus on the time complexity reduction of \sysname based on graph parallel computing \cite{liu2017feasible}.
We also plan to fuse more heterogeneous data \cite{liu2022traffic,liu2018digger} to enrich the recommendation performance while both local and global information are extremely sparse, or consider the condition of privacy protection or dynamical modeling \cite{liu2019coevil}  problems in recommendation systems.

\section{Acknowledgments}

This work is supported by  National Key R\&D Program (2020YFB1406900), National Natural Science Foundation of China (61902308, U21B2018, 62103323, 62161160337, 61822309, 61773310), Initiative Postdocs Supporting Program BX20190275, BX20200270, and China Postdoctoral Science Foundation 2019M663723, 2021M692565, the Fundamental Research Funds for the Central Universities under grant xjh032021058, xxj022019016, xtr022019002 and Shaanxi Province Key Industry Innovation Program (2021ZDLGY01-02).
The authors also would like to thank the reviewers, and as well as Shuai Xiao, for their useful comments and suggestions.

Chao Shen is the corresponding author.

% \input{ref}
\bibliographystyle{ACM-Reference-Format}
\bibliography{ref}

\end{document}